\documentclass[nonacm,sigconf,fleqn,svgnames,screen]{acmart}

\AtBeginDocument{%
	}

\definecolor{ml2rblu}{rgb}{0.02,0.27,0.45}
\definecolor{ml2ryel}{rgb}{0.98,0.72,0.18}
\definecolor{ml2rgrn}{rgb}{0.50,0.71,0.18}
\definecolor{ml2rtrq}{rgb}{0.00,0.57,0.57}
\usepackage{amsthm}
\usepackage{mathtools}
\usepackage{bm}
\usepackage[font=footnotesize]{subfig}
\usepackage{booktabs}
\usepackage{array}
\usepackage[inline]{enumitem}
\usepackage{fixltx2e}
\usepackage{nicefrac}
\usepackage[capitalize]{cleveref}

\usepackage{wasysym}
\usepackage{wrapfig}
\usepackage{xspace}

\usepackage{tikz}
\usetikzlibrary{shapes.geometric, positioning, calc}

\usepackage{algorithm}
\usepackage[noend]{algpseudocode}

\usepackage{hyperref}
\hypersetup{%
	colorlinks=true,
	urlcolor=ml2rtrq,
	linkcolor=ml2rtrq,
	citecolor=ml2rtrq,
	bookmarks=false}

\usepackage{fancyvrb} 
\usepackage{listings}
\lstdefinestyle{pythonstyle}{%
	language=Python,
	tabsize=4,
	backgroundcolor=\color{Gray!5},
	basicstyle=\ttfamily\scriptsize,
	stringstyle=\color{ForestGreen},
	keywordstyle=\color{BlueViolet},
	commentstyle=\itshape\color{DarkRed!90},
	identifierstyle=,
	emphstyle=\color{Blue},
	showstringspaces=false,
	morekeywords={range, len, self, other, lambda, from, import, as, False, True, 
		enumerate, xrange, map, list, set, float, int, min, max, sorted, None},
	fancyvrb=true,
}
\lstnewenvironment{python}[1][]{\lstset{style=pythonstyle,#1}}{}

\lstdefinestyle{cstyle}{%
	language=C,
	tabsize=4,
	backgroundcolor=\color{Gray!10},
	basicstyle=\ttfamily\scriptsize,
	stringstyle=\color{ForestGreen},
	keywordstyle=\color{BlueViolet},
	commentstyle=\itshape\color{DarkRed!90},
	identifierstyle=,
	emphstyle=\color{Blue},
	frame=lines,	
	showstringspaces=false,
	morekeywords={},
	fancyvrb=true,
}
\lstnewenvironment{ccode}[1][]{\lstset{style=cstyle,#1}}{}

\lstdefinestyle{pythonstyletxt}{%
	language=Python,
	tabsize=4,
	basicstyle=\ttfamily\small,
	stringstyle=\color{ForestGreen},
	keywordstyle=\color{BlueViolet},
	commentstyle=\itshape\color{DarkRed!90},
	identifierstyle=,
	emphstyle=\color{Blue},
	xleftmargin=1em,
	showstringspaces=false,
	morekeywords={range, len, self, other, lambda, from, import, as, False, True, 
		enumerate, xrange, map, list, set, float, int, min, max, sorted, with, None},
}
\lstnewenvironment{pythontxt}[1][]{\lstset{style=pythonstyletxt,#1}}{}

\lstdefinestyle{pythonstyletxtsmall}{%
	language=Python,
	tabsize=4,
	basicstyle=\ttfamily\scriptsize,
	stringstyle=\color{ForestGreen},
	keywordstyle=\color{BlueViolet},
	commentstyle=\itshape\color{DarkRed!90},
	identifierstyle=,
	emphstyle=\color{Blue},
	xleftmargin=1em,
	showstringspaces=false,
	morekeywords={range, len, self, other, lambda, from, import, as, False, True, 
		enumerate, xrange, map, list, set, float, int, min, max, sorted, None},
}
\lstnewenvironment{pythontxtsmall}[1][]{\lstset{style=pythonstyletxtsmall,#1}}{}

\newcommand{\putORCID}[1]{
	\authornote{\href{https://orcid.org/#1}{\includegraphics[width=2ex]{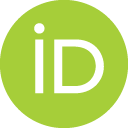}} \href{https://orcid.org/#1}{#1}}
	\orcid{#1}}

\usepackage{tikz}
\usetikzlibrary{shapes.geometric, positioning}
\usepackage{datatool}

\newcommand{\qubolite}{\texttt{qubolite}\xspace}

\newcommand{\bQ}{\bm{Q}}
\newcommand{\bx}{\bm{x}}
\newcommand{\bX}{\bm{X}}
\newcommand{\bz}{\bm{z}}

\newcommand{\T}{\intercal}

\newcommand{\BB}{\lbrace 0,1\rbrace} 
\newcommand{\BP}{\mathbb{P}}
\newcommand{\BR}{\mathbb{R}}
\newcommand{\NP}{\textsf{NP}\xspace}
\newcommand{\SharpP}{\textsf{\#P}\xspace}
\newcommand{\QUBO}{\textsf{QUBO}\xspace}

\newcommand*{\defeq}{\mathrel{\vcenter{\baselineskip0.5ex\lineskiplimit0pt\hbox{\scriptsize.}\hbox{\scriptsize.}}}%
	=}

\DeclarePairedDelimiter\abs\lvert\rvert

\begin{document}

\hypersetup{%
  pdftitle={A Simple QUBO Formulation of Sudoku},
  pdfauthor={Sascha M\"ucke},
  pdfsubject={optimization, qubo, sudoku},
  pdfkeywords={optimization, qubo, sudoku}
}

\title[QUBOLite]{QUBOLite: A lightweigth Python toolkit for QUBO}

\author[S. M\"ucke]{Sascha M\"ucke}
\putORCID{0000-0001-8332-6169}
\affiliation{
  \institution{Lamarr Institute, TU Dortmund University}
  \city{Dortmund}
  \country{Germany}
}
\author[T. Gerlach]{Thore Gerlach}
\putORCID{0000-0001-7726-1848}
\affiliation{
  \institution{Lamarr Institute, University of Bonn}
  \city{Bonn}
  \country{Germany}
}
\author[N. Piatkowski]{Nico Piatkowski}
\putORCID{0000-0002-6334-8042}
\affiliation{
  \institution{Lamarr Institute, Fraunhofer IAIS}
  \city{Sankt Augustin}
  \country{Germany}
}
\author[L. Thei{\ss}inger]{Lukas Thei{\ss}inger}
\affiliation{
  \institution{Lamarr Institute, University of Bonn}
  \city{Bonn}
  \country{Germany}
}

\begin{abstract}
We present \emph{qubolite}, a Python package for the creation, manipulation, analysis, and solution of Quadratic Unconstrained Binary Optimization (\QUBO) instances.
Built as a thin wrapper around NumPy arrays, \emph{qubolite} combines efficient numerical operations with high-level abstractions for tasks ranging from instance generation to preprocessing, analysis, and solving strategies, both exact and approximate.
The package includes implementations of the QPRO+ algorithm by Glover et al. for identifying strong persistencies, dynamic range reduction heuristics, and an expressive system for partial assignments (clamping) enabling implicit variable assignment.
The package is available on GitHub and the official Python package repository.
\end{abstract}

\maketitle

\section{Introduction}
\label{sec:intro}

Quadratic Unconstrained Binary Optimization (\QUBO) is a versatile class of combinatorial optimization problems, defined as the problem of finding a binary vector $\bz^*\in\BB^n$ that minimizes the quadratic pseudo-boolean function $E_{\bQ}(\bz)\defeq\bz^\T\bQ\bz$, i.e., \begin{equation*}
	\forall\bz\in\BB^n: ~E_{\bQ}(\bz^*)\leq E_{\bQ}(\bz),
\end{equation*}%
for a given upper-triangular or symmetric \emph{weight matrix} $\bQ\in\BR^{n\times n}$.
The function $E_{\bQ}$ is sometimes referred to as the \emph{energy function}.
Some authors prefer to define \QUBO as the problem of \emph{maximizing} $E_{\bQ}$, but here we always assume minimization.

In its general form, \QUBO is strongly \NP-hard~\cite{pardalos.jha.1992b,cela.punnen.2022a}, as its solution space grows exponentially in $n$.
Its value lies in its adaptability to a wide range of combinatorial optimization problems, ranging from economics~\cite{laughhunn.1970a,hammer.shlifer.1971a} over satisfiability~\cite{kochenberger.etal.2005a}, resource allocation and routing problems~\cite{neukart.etal.2017a,stollenwerk.etal.2019a} to machine learning~\cite{bauckhage.etal.2018a,muecke.etal.2019a,date.etal.2020a,muecke.etal.2023a}, to name just a few.
Its structural simplicity has made it a popular target problem for special-purpose hardware solvers~\cite{matsubara.etal.2017a,muecke.etal.2019b}.
Notably, \QUBO can be mapped to an Ising model~\cite{brush.1967a} and solved through quantum annealing, which exploits quantum tunneling effects~\cite{kadowaki.nishimori.1998a} and has lead to renewed interest in this problem class.

\medskip
In this paper we present \qubolite, a versatile toolbox for creating, analyzing, manipulating and solving \QUBO instances.

\subsection{Design Principles}

The \qubolite package provides a \verb|qubo| class, which is a shallow wrapper around \verb|numpy| arrays providing basic convenience functions for working with \QUBO instances.
By design, our package does not provide much guidance for \emph{formulating} \QUBO problems, e.g., by computing the weight matrix from other problem formulations or from given sets of constraints (although it comes with the submodule \verb|embedding| providing a few common problem embeddings).
Instead, \qubolite focuses on analyzing, preprocessing and solving existing \QUBO instances on a numerical level, using the weight matrix $\bQ$ as the unique characterization of a \QUBO problem.
To this end, it provides a wide range of useful classes and methods, prioritizing clarity, usability and intuition.

\begin{figure}
	\centering
	\includegraphics[width=.5\columnwidth]{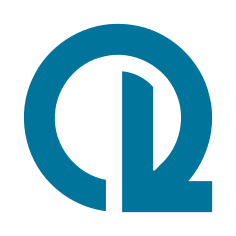}
	\caption*{QUBOLite logo}
\end{figure}

As an example, the following code \begin{enumerate}[label=(\roman*)]
	\item samples a random \QUBO instance of size $n=16$ with Gaussian weights,
	\item clamps a few variables (see \cref{sec:pa}),
	\item computes its dynamic range,
	\item performs a dynamic range reduction to improve the solution quality on quantum annealers (see \cref{sec:dr}),
	\item solves the resulting \QUBO instance by brute force (see \cref{sec:solving}), and
	\item re-inserts the implicit clamped variables into the solution.\end{enumerate}
\begin{python}
from qubolite               import qubo
from qubolite.assignment    import partial_assignment
from qubolite.preprocessing import reduce_dynamic_range
from qubolite.solving       import brute_force

Q = qubo.random(16, distr='normal', density=0.8) # (i)

clamp = partial_assignment('x1=0; x5=!x4', n=16)
Q_, const = clamp.apply(Q) # (ii)
print(Q_.dynamic_range()) # (iii)
# 19.523642419161717

Q_reduced = reduce_dynamic_range(Q_) # (iv)
print(Q_reduced.dynamic_range())
# 14.302202421444324

solution = brute_force(Q_reduced) # (v)
x_ = solution.x
print(x_)
# [1. 1. 0. 0. 1. 0. 1. 1. 1. 0. 1. 1. 1. 1.]

x = clamp.expand(x_) # (vi)
print(x)
# [1. 0. 1. 0. 0. 1. 1. 0. 1. 1. 1. 0. 1. 1. 1. 1.]
\end{python}

The following sections highlight some notable features of the \qubolite package.
A full documentation can be found at \begin{center}
	\url{https://smuecke.de/qubolite/api.html}.
\end{center}

\section{Working with \QUBO Instances}

The \qubolite module provides the \verb|qubo| class, which wraps around NumPy arrays of shape \verb|(n, n)| and treats them as the weight matrix of a \QUBO instance.
\begin{python}
import numpy as np
from qubolite import qubo

weights = np.random.normal(size=(16, 16))
Q = qubo(weights)

print(Q.n) # size (= number of variables)
# 16

print(Q.dynamic_range())
# 15.179273963592772

# convert to Ising model
h, J, const = Q.to_ising()
\end{python}

To evaluate the energy function, a \verb|qubo| object can be called just like a function.
Energy evaluation is fully vectorized, allowing to evaluate many binary vectors at once.
\begin{python}
from qubolite.bitvec import all_bitvectors_array, from_string

x = from_string('0110110111110111')
print(Q(x))
# 8.701159165768138

X = all_bitvectors_array(16)
print(X.shape)
# (65536, 16)

E = Q(X)
print(E.shape)
# (65536,)
\end{python}

\QUBO instances can be saved and loaded to and from disk using a custom file format that automatically chooses the most efficient data representation, depending on the weight matrix's density (i.e., the number of non-zero weights).

By default, \qubolite always uses the upper triangular form of \QUBO weight matrices and converts non-triangular matrices passed to the \verb|qubo| class to this form by applying \begin{equation*}
	Q_{ij}\mapsto\begin{cases}Q_{ij}+Q_{ji}&\text{if }i<j,\\Q_{ii}&\text{if }i=j,\\ 0&\text{otherwise.}\end{cases}
\end{equation*}
Consequently, for any \verb|qubo| object \verb|Q|, the parameter matrix \verb|Q.m| is upper triangular.
To convert to the symmetrical form:
\begin{python}
Q_symm = (Q.m + Q.m.T)/2
\end{python}

\subsection{Discrete Derivatives}

The class also provides methods for easily computing the discrete derivative of a \QUBO instance~\cite{boros.etal.2007a} w.r.t. some binary vector $\bx\in\BB^n$, i.e., the change in energy when flipping bit $x_i$ for every $i=1,\dots,n$.
\begin{python}
dEdx = Q.dx(x)
print(dEdx)
# array([ 0.41239916, -1.30958185,  2.98111447,  0.89453998,
#        -4.76395142,  0.4630076 ,  0.1078992 , -3.10359708, 
#         0.31572925, -7.87009611,  6.04688876,  2.47012002,
#        -0.49744586,  0.92405999,  1.27467536,  0.76755653])

dEdX = Q.dx(X)
print(dEdX.shape)
# (65565, 16)
\end{python}

The 2\textsuperscript{nd} discrete derivative is a matrix $\bm\Delta\in\BR^{n\times n}$ with $\Delta_{ij}$ being the change in energy when bits $i$ and $j$ are flipped at the same time.
The diagonal of $\bm\Delta$ is the same as the first discrete derivative.
\begin{python}
dEdx2 = Q.dx2(x)
print(dEdx2.shape)
# (16, 16)
\end{python}

\subsection{Gibbs Distribution}

Every \QUBO instance can be interpreted as a \emph{Gibbs distribution} (or \emph{Boltzmann distribution}) over the set $\BB^n$.
The probability of a random variable $\bX$ taking the value $\bx$ is given by \begin{align*}
	\BP[\bX=\bx\mid\bQ,\beta] &=P(\bx;\bQ,\beta)=\frac{1}{Z_{\bQ,\beta}}e^{-\beta E_{\bQ}(\bx)},\\
	\text{where }Z_{\bQ,\beta}&=\hspace{-1em}\sum_{\bx'\in\BB^n}\hspace{-1em}e^{-\beta E_{\bQ}(\bx)}.
\end{align*}
The value $\beta>0$ is the inverse temperature, a hyperparameter controlling the entropy of the probability distribution.
$Z_{\bQ,\beta}$ is the \emph{partition function}, acting as a normalization constant so that $\sum_{\bx'\in\BB^n}P(\bx')=1$.
Computing this constant is \SharpP-complete in general~\cite{valiant.1979a}.

Objects of the \verb|qubo| class have some builtin functionality to work with this probabilistic interpration of \QUBO:
\begin{python}
Q = qubo.random(16)

# compute the (log) partition function
logZ = Q.partition_function(log=True) # beta = 1, by default
print(logZ)
# 20.09671604952432

# compute pairwise marginal probabilities of binary variables
M = Q.pairwise_marginals()
print(f'P[x0=1 & x4=1] = {M[0, 4]}')
# P[x0=1 & x4=1] = 0.6579248732567058

# compute full probability vector for all binary vectors
P = Q.probabilities()
print(P.shape)
# (65536,)
\end{python}
Due to the high runtime/memory complexity of these operations, they are only feasible for small \QUBO sizes.

In addition to these functions, the \verb|sampling| submodule provides a class \verb|BinarySample| representing collections of binary vectors, with useful methods relating to empirical probability distributions over $\BB^n$, such as empirical probability, subsampling, computing the sufficient statistic for pairwise models, and computing the Hellinger distance.

\section{Preprocessing}

In addition to storing and analyzing \QUBO instances, \qubolite has a submodule \verb|preprocessing| containing methods to modify \verb|qubo| objects in order to \begin{enumerate*}[label=(\roman*)]\item fix some of their variables or \item reduce their dynamic range for better performance on solvers with limited precision.\end{enumerate*}

As a precursor, we first introduce \emph{partial assignments}, an important tool for working with implicit variable assignments.

\subsection{Partial Assignments}
\label{sec:pa}

The submodule \verb|assignment| contains the class \verb|partial_assignment|, which can be used to encode the act of fixing certain variables to constants, or to tie the value of one variable to the value of another.
This procedure is sometimes known as \emph{clamping} in literature~\cite{booth.etal.2017a,muecke.2024a}.

There are three ways to instantiate a \verb|partial_assignment| object, namely using \begin{enumerate*}[label=(\roman*)]\item assignment expressions, \item bit vector expression, and \item dictionaries,\end{enumerate*} allowing for flexible and intuitive usage:
\begin{python}
from qubolite.assignment import partial_assignment

# instanstiate using an assignment expression
PA = partial_assignment(
    'x0, x3 = 0; x7 = 1; x12 = x8; x13 != x9', n=16)

# instantiate using a bit vector expression
PA = partial_assignment.from_expression('**00**[1]*1[!4]1')

# instantiate from a dictionary
PA = partial_assignment.from_dict({0: 1, 1: 1, 5: 0}, n=10)
\end{python}

Once a \verb|partial_assignment| object is defined, it can be used to \begin{enumerate*}[label=(\roman*)]\item make variables in a \QUBO instance implicit, reducing its size in the process, \item re-insert implicit variables into bit vectors, and \item generate all bit vectors matching its pattern,\end{enumerate*} among other things.
\begin{python}
import numpy as np
from qubolite.bitvec import random
	
# apply partial assignment to QUBO instance Q
Q_, const = PA.apply(Q)
# returns a smaller QUBO instance and a constant
# term that can be used to recover the energy value

x_ = random(Q_.n)
E_ = Q_(x_)

x = PA.expand(x_)
E = Q(x)

print(np.isclose(E, E_ + const))
# True
\end{python}

\subsection{QPRO+}
Glover et al.~\cite{glover.etal.2018a} derive a number of conditions under which a specific value of a binary variable in a \QUBO instance must be optimal, or which variable pairs must have the same or opposite values.
They propose the QPRO+ algorithm to find those assignments.
The \verb|preprocessing| submodule of \qubolite contains an implementation of this algorithm, which takes a \verb|qubo| object as input and returns a partial assignment that can be applied to reduce the size of the original \QUBO instance:

\begin{python}
from qubolite.preprocessing import qpro_plus

# sample random QUBO instance
Q = qubo.random(20, density=0.3)

# run QPRO+ algorithm, which returns a partial assignment
PA = qpro_plus(Q)
print(PA)
# x5, x8, x11, x13 = 0; x0, x3, x6, x9, x15, x16 = 1; x17 != x1

# apply partial assignment to reduce QUBO size
Q_, const = PA.apply(Q)
print(Q_.n)
# 9
\end{python}

\subsection{Dynamic Range Reduction}
\label{sec:dr}

Recent work by M\"ucke et al. demonstrated that the dynamic range of a \QUBO instance or Ising model has a significant impact on the solution quality of quantum annealers, which represent the weights $Q_{ij}$ with low floating point precision~\cite{muecke.etal.2024a}.
The dynamic range of a \QUBO instance is defined as the logarithmic ratio between the largest and smallest difference between unique weights:
\begin{align*}
	\operatorname{DR}(\bQ)&=\log_2\biggl(\frac{\max D(\bQ)}{\min D(\bQ)}\biggr)\\
	\text{where }D(\bQ) &= \lbrace\abs{Q_{ij}-Q_{kl}}: ~Q_{ij}\neq Q_{kl}\rbrace
\end{align*}
The authors devise a heuristic algorithm to reduce the dynamic range of any \QUBO instance by exploiting upper and lower bounds on the minimal energy on subspaces of $\BB^n$.
An implementation of this algorithm can be found in the \verb|preprocessing| submodule:
\begin{python}
from qubolite.preprocessing import reduce_dynamic_range
	
Q = qubo.random(32, density=0.25)
print(Q.dynamic_range())
# 17.02497209387645

Q_ = reduce_dynamic_range(Q)
print(Q_.dynamic_range())
# 8.398588438912235
\end{python}

This heuristic algorithm works best for small instances or instances with low density.

\section{Solving}
\label{sec:solving}

The submodule \verb|solving| provides a few common methods for solving \QUBO instances approximately, such as Simulated Annealing~\cite{kirkpatrick.etal.1983a} and different local search heuristics.
However, the focus of \qubolite is not to provide highly efficient implementations of \QUBO solvers.

A notable feature is our fast parallel C implementation of a brute-force solver, which uses Gray codes to compute the energy of all bit vectors more efficiently~\cite{muecke.2023a}.
While brute force is not a feasible strategy for larger \QUBO instances beyond around 30 variables, this implementation is useful for testing and research purposes.

\section{Usage}

You can install \qubolite either from PyPI or directly from the repository on GitHub:
\begin{verbatim}
	pip install qubolite
	pip install git+https://github.com/smuecke/qubolite
\end{verbatim}
It requires at least Python version 3.8.
For the most up-to-date features, you can install the package directly from the \verb|dev| branch, at the risk of encountering a few unstable or undocumented features:
\begin{verbatim}
	pip install git+https://github.com/smuecke/qubolite@dev
\end{verbatim}

The \qubolite package may be used freely for research and private purposes.
If you use this package in your research, please cite this paper as a reference.
For other use, e.g., in commercial applications, please contact: \begin{center}
	\texttt{sascha.muecke@tu-dortmund.de}
\end{center}

\section*{Acknowledgments} 

This research has been funded by the Federal Ministry of Education and Research of Germany and the state of North-Rhine Westphalia as part of the Lamarr Institute for Machine Learning and Artificial Intelligence.

\end{document}